\documentclass[aps,prd,twocolumn,superscriptaddress]{revtex4}

\usepackage{graphicx}

\begin{document}

\newcommand{\kthreep}{\mbox{$K_{L}\rightarrow 3\pi^0$}}
\newcommand{\ptoee}{\mbox{$\pi^0\rightarrow e^+e^-$}}
\newcommand{\ptoeeg}{\mbox{$\pi^0\rightarrow e^+e^-\gamma$}}


\title{Measurement of the rare decay $\pi^0\rightarrow e^+e^-$\\
\normalsize
}




%

\newcommand{\UAz}{University of Arizona, Tucson, Arizona 85721}
\newcommand{\UCLA}{University of California at Los Angeles, Los Angeles,
                    California 90095}
\newcommand{\Campinas}{Universidade Estadual de Campinas, Campinas,
                       Brazil 13083-970}
\newcommand{\EFI}{The Enrico Fermi Institute, The University of Chicago,
                  Chicago, Illinois 60637}
\newcommand{\UB}{University of Colorado, Boulder, Colorado 80309}
\newcommand{\ELM}{Elmhurst College, Elmhurst, Illinois 60126}
\newcommand{\FNAL}{Fermi National Accelerator Laboratory,
                   Batavia, Illinois 60510}
\newcommand{\Osaka}{Osaka University, Toyonaka, Osaka 560-0043 Japan}
\newcommand{\Rice}{Rice University, Houston, Texas 77005}
\newcommand{\SaoPaolo}{Universidade de S\~ao Paulo, S\~ao Paulo, Brazil 05315-970}
\newcommand{\UVa}{The Department of Physics and Institute of Nuclear
and
                  Particle Physics, University of Virginia,
                  Charlottesville, Virginia 22901}
\newcommand{\UW}{University of Wisconsin, Madison, Wisconsin 53706}

\affiliation{\UAz}
\affiliation{\UCLA}
\affiliation{\Campinas}
\affiliation{\EFI}
\affiliation{\UB}
\affiliation{\ELM}
\affiliation{\FNAL}
\affiliation{\Osaka}
\affiliation{\Rice}
\affiliation{\SaoPaolo}
\affiliation{\UVa}
\affiliation{\UW}

\author{E.~Abouzaid}  \affiliation{\EFI}
\author{M.~Arenton}       \affiliation{\UVa}
\author{A.~R.~Barker}      \altaffiliation{Deceased.} \affiliation{\UB}
\author{L.~Bellantoni}    \affiliation{\FNAL}
\author{A.~Bellavance}    \affiliation{\Rice} 
\author{E.~Blucher}       \affiliation{\EFI}
\author{G.~J.~Bock}        \affiliation{\FNAL}
\author{E.~Cheu}          \affiliation{\UAz}
\author{R.~Coleman}       \affiliation{\FNAL}
\author{M.~D.~Corcoran}    \affiliation{\Rice}
\author{B.~Cox}           \affiliation{\UVa}
\author{A.~R.~Erwin}       \affiliation{\UW}
\author{C.~O.~Escobar}     \affiliation{\Campinas}  
\author{A.~Glazov}        \affiliation{\EFI}
\author{A.~Golossanov}    \affiliation{\UVa} 
\author{R.~A.~Gomes}       \affiliation{\Campinas}
\author{P. Gouffon}       \affiliation{\SaoPaolo}
\author{K.~Hanagaki}      \affiliation{\Osaka} 
\author{Y.~B.~Hsiung}      \affiliation{\FNAL}
\author{H.~Huang}         \affiliation{\UB}    
\author{D.~A.~Jensen}      \affiliation{\FNAL}
\author{R.~Kessler}       \affiliation{\EFI}
\author{K.~Kotera}  \affiliation{\Osaka}
\author{A.~Ledovskoy}     \affiliation{\UVa}
\author{P.~L.~McBride}     \affiliation{\FNAL}

\author{E.~Monnier}
   \altaffiliation[Permanent address ]{C.P.P. Marseille/C.N.R.S., France}
   \affiliation{\EFI}  

\author{K.~S.~Nelson}     \affiliation{\UVa}  
\author{H.~Nguyen}       \affiliation{\FNAL}
\author{R.~Niclasen}     \affiliation{\UB}
\author{D.~G.~Phillips~II} \affiliation{\UVa}
\author{H.~Ping}         \affiliation{\UW}  
\author{X.~R.~Qi}         \affiliation{\FNAL} 
\author{E.~J.~Ramberg}    \affiliation{\FNAL}
\author{R.~E.~Ray}        \affiliation{\FNAL}
\author{M.~Ronquest}     \affiliation{\UVa}
\author{E.~Santos}       \affiliation{\SaoPaolo}
\author{J.~Shields}      \affiliation{\UVa} 
\author{W.~Slater}       \affiliation{\UCLA}
\author{D.~Smith}        \affiliation{\UVa}
\author{N.~Solomey}      \affiliation{\EFI}
\author{E.~C.~Swallow}    \affiliation{\EFI}\affiliation{\ELM}
\author{P.~A.~Toale}      \affiliation{\UB}
\author{R.~Tschirhart}   \affiliation{\FNAL}
\author{C.~Velissaris}   \affiliation{\UW}  
\author{Y.~W.~Wah}        \affiliation{\EFI}
\author{J.~Wang}         \affiliation{\UAz}
\author{H.~B.~White}      \affiliation{\FNAL}
\author{J.~Whitmore}     \affiliation{\FNAL}
\author{M.~J.~Wilking}      \affiliation{\UB}
\author{B.~Winstein}     \affiliation{\EFI}
\author{R.~Winston}      \affiliation{\EFI}
\author{E.~T.~Worcester}  \affiliation{\EFI}
\author{M.~Worcester}    \affiliation{\EFI}
\author{T.~Yamanaka}     \affiliation{\Osaka}
\author{E.~D.~Zimmerman} \altaffiliation{To whom correspondence should be
addressed. Electronic address: {\tt edz@colorado.edu}} \affiliation{\UB}
\author{R.~F.~Zukanovich} \affiliation{\SaoPaolo}

\date{\today}

\begin{abstract}

The branching ratio of the rare decay $\pi^0\rightarrow e^+e^-$ has
been measured precisely, using the complete data set from the KTeV
E799-II experiment at Fermilab.  We observe 794 candidate \ptoee\ 
events using \kthreep\  as a source of tagged $\pi^0$s. The expected
background is $52.7\pm 11.2$ events, predominantly from high $e^+e^-$
mass \ptoeeg\  decays.  We have measured B($\pi^0\rightarrow e^+e^-,
~(m_{e^+e^-}/m_{\pi^0})^2> 0.95$) = $(6.44\pm 0.25_{\rm stat}\pm 0.22_{\rm syst})\times
10^{-8}$, which is above the unitary bound from $\pi^0\rightarrow\gamma\gamma$
and within the range of theoretical expectations from the standard model.

\end{abstract}


\maketitle



\section{\label{intro} INTRODUCTION}

In this report we present a new measurement of the \ptoee\ branching
ratio using a larger data set from KTeV-E799 at Fermilab. This result 
used all data taken in the two runs of the experiment (1997 and
1999-2000). It supersedes the previously published measurement
\cite{zimmerman} from KTeV-E799, which used only the 1997 data.
The basic measurement technique of using \kthreep\ as a source
of tagged $\pi^0$ decays is adapted from the previous analysis. 

\subsection{The Decay \ptoee }

The rare decay \ptoee\ proceeds, to lowest order, in a one-loop process
via a two-photon intermediate state. The decay rate was first
predicted by Drell \cite{drell} and has since received considerable
attention both theoretically and experimentally. Relative to the
$\pi^0\rightarrow \gamma\gamma$ rate, it is suppressed by two powers
of $\alpha$ and is further suppressed by $2(m_e/m_{\pi^0})^2$ due to
the approximate helicity conservation of the interaction.  The lowest
order contribution has been calculated exactly in terms of a form
factor \cite{bergstrom_one}, and lowest order radiative corrections
have been calculated \cite{bergstrom_two}. The contribution to the
rate from on-shell photons is model independent and can be calculated
exactly to form a lower ``unitary bound'' \cite{berman} on the branching ratio,
B(\ptoee)$\geq 4.69\times 10^{-8}$, neglecting radiative corrections.  

The primary interest in the decay rate is the excess above the unitary
bound, as this is the contribution from virtual photons.  Attempts to
model the form factor and make predictions for the off-shell photon
contribution have been made, most successfully using vector meson
dominance (VMD) and chiral perturbation theory 
($\chi$PT) approaches \cite{ametller,savage,gomez,knecht}.
A new measurement is significant for $\chi$PT, where
\ptoee\ represents a tight experimental constraint on
calculations.  It is of particular interest because \ptoee\ is the
best-measured decay of a pseudoscalar meson to a lepton pair and has
no significant contributions from short-distance physics.  Constraints
on $\chi$PT from \ptoee\ can be used to improve predictions for other
$P^0\rightarrow l^+l^-$ decays, including $\eta\rightarrow\mu^+\mu^-$
and the long-distance contribution to
$K^0_L\rightarrow\mu^+\mu^-$. The smaller short-distance contribution
to $K^0_L\rightarrow\mu^+\mu^-$ is dominated by a top-quark loop and
thus is a potential source of information on $|V_{td}|$ if the
long-distance contribution can be subtracted successfully.

Earlier interest in \ptoee\ was due to experimental
indications \cite{fischer,frank} that the decay rate could be
substantially higher than predicted, indicating possible new
physics. Later experiments
\cite{niebuhr,deshpande,mcfarland} obtained results more
consistent with the standard model predictions, and the most recent
result from KTeV-E799 \cite{zimmerman} provided a precise
measurement of the branching ratio falling entirely within the
standard model prediction.

\section{THE KTeV-E799 EXPERIMENT}

\begin{figure*}
\includegraphics[width=6in]{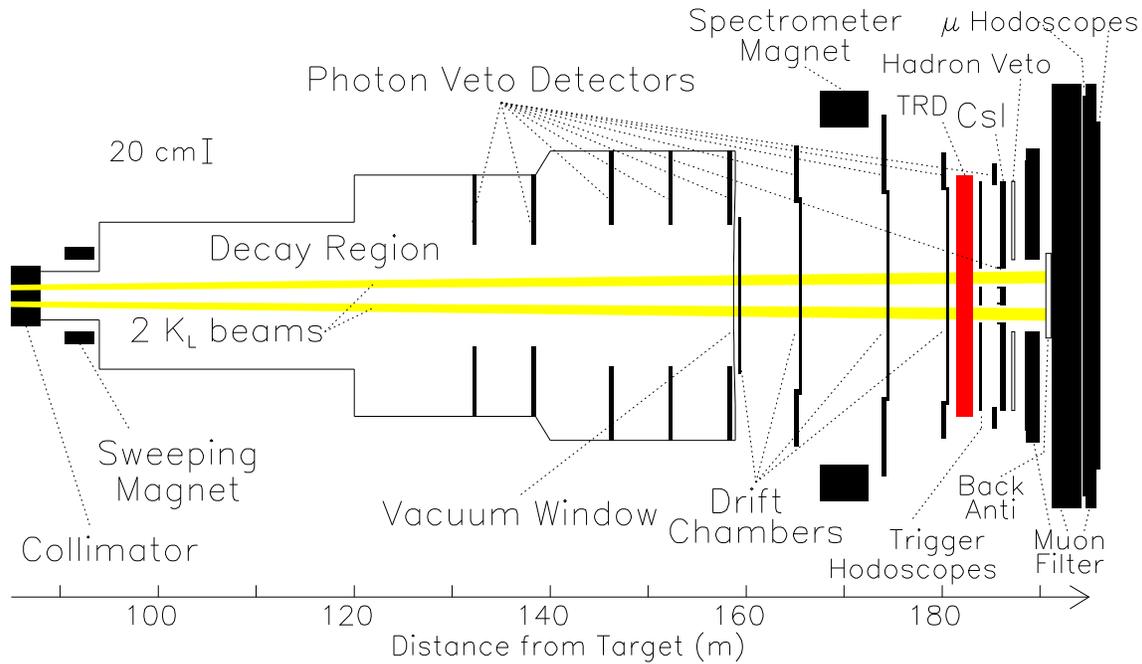}
\caption{\label{spectrometer} The KTeV spectrometer in E799-II configuration.
}
\end{figure*}

The KTeV facility (Figure~\ref{spectrometer}) at Fermilab was a
general-purpose neutral kaon beam and decay spectrometer.  It was
operated for two experiments: E832, which used an active regenerator
to produce a $K_S$ flux for measuring $\Re
(\epsilon^\prime/\epsilon)$, and E799-II, which had a higher intensity
$K_L$ flux and performed searches and measurements for a variety of
rare $K_L$ decays. The analysis described here uses E799-II data.

\subsection{The kaon beam}

At KTeV, 800 GeV protons hit a BeO target and produced two nearly
parallel neutral beams that were defined by sweeper magnets and
collimators. A vacuum decay volume was located from 94 to 158 meters
downstream of the target.  In this region, the beams consisted of
$K^0_L$ and neutrons, with small numbers of shorter-lived neutral
hyperons and $K^0_S$ remaining.  The $K^0_L$ energies in this region
ranged from 20--200 GeV.  The decay region ended at a Mylar-Kevlar
vacuum window which was followed by a charged particle spectrometer.

\subsection{The charged spectrometer}

Charged tracks were detected by four drift chambers separated by 6 m,
9 m, and 6 m. A momentum analysis dipole magnet sat between the second
and third chambers. The field integral from the magnet was 205 MeV/$c$
in the 1997 run period and 150 MeV/$c$ in 1999. The momentum resolution
of the spectrometer in the range of interest was 0.9\%. A set of
transition radiation detectors (TRDs) was placed after the the last
drift chamber. This detector provided particle identification used to
distinguish electrons from pions but was not needed in this analysis
because there were no significant non-electron backgrounds.  
Following the TRDs there was a segmented array of scintillator planes for
fast triggering on events with charged particles. 

\subsection{Photon and muon detection}

The final detector for electromagnetic particles was a calorimeter consisting
of 3100 pure CsI crystals. The crystal blocks were arranged in a $1.8
\times 1.8$~m$^2$ square array with two $15\times 15$~cm$^2$ holes
near the middle for the neutral beams to pass through. The crystals
were 27 radiation lengths deep, which contained nearly all
electromagnetic showers.  The energy resolution for electromagnetic
particles was $(0.45 \oplus 2.0/\sqrt{E})\%$, where $E$ is the energy
in GeV and the addition is in quadrature.  The perimeters of the vacuum
decay region, spectrometer, and calorimeter were instrumented with a
total of nine lead-scintillator photon veto counters to reject particles
escaping the detector at high angles. Two vetoes were also used around
the edges of the two beam holes in the calorimeter.  Downstream of the
calorimeter, a 15~cm lead wall showered remaining hadrons and the
showers were detected by a scintillator plane in order to reject
events with hadrons in the final state. Behind an additional 4 m of
steel was a muon veto system, which was used to detect decays with
muons in the final state.  For a more complete discussion of the KTeV
detector see Ref.~\cite{graham}.

\section{DATA COLLECTION AND ANALYSIS}

The \ptoee\ branching ratio was found by normalizing signal candidates
to 
$\pi^0\rightarrow e^+e^-\gamma$ (Dalitz decays) 
with $m_{e^+e^-}>65$ MeV/$c^2$.  Both samples were from \kthreep\
decays where the other two $\pi^0$'s in the event decayed to
$\gamma\gamma$.  This normalization mode was selected because its
final state particles and kinematics are similar to the signal,
allowing many detector response systematic effects to cancel.

\subsection{Trigger requirements}

The trigger for both signal and normalization required activity in the
chambers consistent with two tracks, plus total energy in the
calorimeter above 25 GeV and at least four separate energy clusters in
the calorimeter where at least one crystal in the cluster had more than 1 GeV
of energy. The trigger also required no significant energy in either
the photon veto counters or the hadron anti.
Signal and normalization candidates were collected, reconstructed, and
analyzed in parallel.  

\subsection{Radiative corrections}

The presence of internal bremsstrahlung off the electrons in \ptoee\
complicates the analysis, because the final state contains the same
particles as the tree-level Dalitz decay \ptoeeg\ (though the two
decays generally populate different regions of phase space). The
signal must therefore be defined as a region where radiation is soft
and where there is little contribution from the Dalitz
decay. Following the conventions of Ref.~\cite{mcfarland,zimmerman}, we
defined the signal by requiring Dalitz
$x_D\equiv(m_{e^+e^-}/m_{\pi^0})^2>0.95$, considering the rest of the
spectrum as background. This definition left very little intrinsic
background from the Dalitz decays while including 89\% of the \ptoee\
bremsstrahlung spectrum. Also, in this region the quantum mechanical
interference between the two
modes is negligible \cite{bergstrom_two}. In addition to the inner
bremsstrahlung diagram, a virtual photon correction suppresses the total
\ptoee\ decay rate by 3.4\%.  Both effects must be accounted for in 
comparing the measured decay rate with theoretical models that neglect
radiation. 

The experimentally measured quantity was the ratio:
\begin{equation}
\frac
{\Gamma(\pi^0\rightarrow e^+e^-,~~x_D>0.95)}
{\Gamma(\pi^0\rightarrow e^+e^-\gamma,~~x_D>0.232)}
\end{equation}
where the \ptoeeg\ rate is inclusive of $\pi^0 \rightarrow e^+e^-\gamma\gamma$
as calculated from lowest-order radiative corrections \cite{mikaelian}. 

\subsection{Reconstruction and event selection}

For both modes the full \kthreep\ decay chain was reconstructed. 
Signal events had 6 electromagnetic clusters and 2 oppositely charged
tracks, while the normalization had 7 clusters and 2 tracks. The
tracks in both modes also had to be electron candidates, defined to be
the case when a track of momentum $p$ pointed to a calorimeter cluster
of energy $E$ and $|E/p-1|\leq 0.08$. The total energy in the
calorimeter was required to be above 35 GeV and each cluster energy
above 1.75 GeV.

\subsection{Photon and vertex reconstruction}

Clusters with no tracks pointing to them were assumed to be photons
coming from $\pi^0$ decays. For \ptoee\ candidates, the four photons
could be assigned in three possible pairing combinations, while for
$\pi^0\rightarrow e^+e^-\gamma$ candidates there were 15 pairing
combinations for the five photons. The best pairing was found using the
following procedure: For each pair of photons the distance $d$ from
the calorimeter to the decay vertex was calculated assuming the pair
originated from a $\pi^0\rightarrow\gamma\gamma$ decay: $d =
r_{12}\sqrt{E_1E_2}/m_{\pi^0}$, where $r_{12}$ was the
distance between the two photon clusters and $E_1$ and $E_2$ were the
cluster energies. The $z$-position of the decay vertex was then
$z=z_{\text{CsI}} - d$. A pairing $\chi^2$ was calculated 
for the hypothesis that the two decay positions ($z_1$ and $z_2$)
coincided with each other and with the decay vertex obtained from the
electron tracks ($z_{ee}$):
\begin{equation}
\chi^2 = \frac{(z_1-\bar{z})^2}{\sigma^2(z_1)} + \frac{(z_2-\bar{z})^2}{\sigma^2(z_2)} + \frac{(z_{ee}-\bar{z})^2}{\sigma^2(z_{ee})}.
\end{equation}
For each pairing case, the mean decay position $\bar{z}$ was found by
minimizing the $\chi^2$. The pairing with the smallest minimum
$\chi^2$ was selected and the obtained decay vertex $z$-position,
$\bar{z}$, was then used to reconstruct particle trajectories. This
decay vertex calculation combined information from the calorimeter and
drift chambers to optimize the overall resolution on the vertex
position. The vertex was required to be $96\leq\bar z\leq 158$~m
downstream of the target, removng events near the ends of the decay
region.

\subsection{Final sample selection}

For \ptoee\ candidates, the reconstructed kaon mass was required to be
between 490--510~MeV/$c^2$.  For normalization \ptoeeg\ candidates,
where backgrounds were low and event reconstruction was poorer due to
the additional pairing ambiguity, the allowed interval was
475--525~MeV/$c^2$. The total reconstructed momentum transverse to the
incident kaon direction, defined as the line between the center of the
target and the decay vertex, was required to be $p_{\bot}^2<10^{-3}$
GeV$^2$/$c^2$.  For the normalization sample the reconstructed Dalitz
decay mass $m_{e^+e^-\gamma}$ was required to be in the interval
100--200~MeV/$c^2$, with an additional requirement that the
reconstructed electron pair mass $m_{e^+e^-}$ be greater than
70~MeV/$c^2$. This last requirement removed resolution effects near
the 65 MeV/$c^2$ cutoff.

\begin{figure}
\includegraphics[width=3.4in]{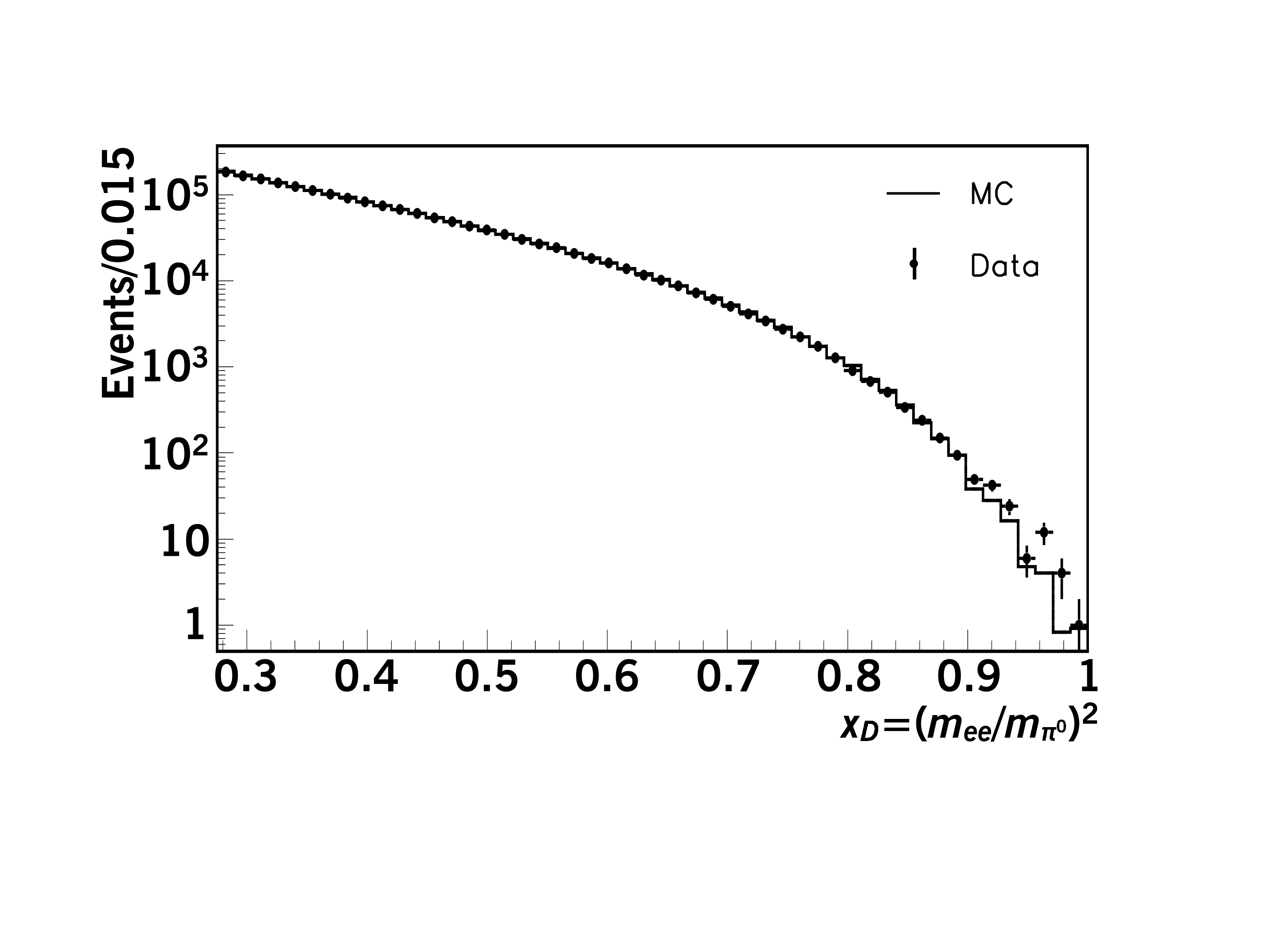}
\caption{\label{xdist}
Dalitz $x_D \equiv (m_{e^+e^-}/m_{\pi^0})$ for fully-reconstructed
Dalitz decay candidates after normalization mode analysis cuts. Monte Carlo
prediction using the world average form factor slope parameter is overlaid.}
\end{figure}

A detailed description of the detector and beamline was implemented in
a Monte Carlo (MC) simulation, which was used to study detector
geometry, acceptance, and backgrounds. The decay simulation included
$\mathcal{O}(\alpha)$ radiative corrections to \ptoee\ based on the
work of Bergstr\"om \cite{bergstrom_two}, while for $\pi^0\rightarrow
e^+e^-\gamma$, the world average form factor slope \cite{pdg} and
radiative corrections to order $\mathcal{O}(\alpha^2)$
\cite{mikaelian} were used.  Figure~\ref{xdist} shows the distribution
of $x_D$ for the normalization Dalitz-decay
sample in both data and MC.

Beyond the basic reconstruction requirements above, additional cuts
were needed to remove backgrounds to \ptoee .  The full reconstruction of the
$K_L$ decay chain removed all significant backgrounds except those
originating from \kthreep\ decays.  One category of backgrounds 
was \kthreep\ decays with four
electrons in the final state, where two were lost and the remaining
two mimicked the \ptoee\ decay. Low-energy electrons could be swept
out of the fiducial region by the analysis magnet, never making a
complete track.  One major source of this background was $\kthreep
\rightarrow (e^+e^-\gamma) + (e^+e^-\gamma) + (\gamma\gamma)$, where the
photons from the Dalitz decays were accidentally reconstructed as a
$\pi^0\rightarrow \gamma\gamma$ decay. Another source was the rarer
decay $\pi^0\rightarrow e^+e^-e^+e^-$.  Finally, photons from $\pi^0$
decays could convert to $e^+e^-$ pairs in the vacuum window just
upstream of the chambers. Events with two of these conversions, or one in
combination with a Dalitz decay, also contributed to four-track
background.

Backgrounds in which the two electrons came from different $\pi^0$s were
reduced by a requirement on the pairing $\chi^2$ defined above: a cut of
$\chi^2<20$ was used in both the signal and the normalization mode.
To reduce the four-electron backgrounds further, a cut on evidence for
extra in-time activity in the second drift chamber was made.  Removing
events with in-time activity in the second drift chamber more than
0.5~cm away from any reconstructed track reduced the four-track
backgrounds to 0.7\% of the expected signal.  The effect of this cut
on the signal (and any backgrounds without extra charged particles)
was an overall reduction of 7.7\% in acceptance.  Both of these
background cuts were also used in the Dalitz normalization sample in
order to cancel systematic effects associated with modeling the cut
efficiency.

\subsection{Background and systematic error estimation}

After all cuts were applied, the largest remaining background came
from high $m_{e^+e^-}$ Dalitz decays where the Dalitz photon was lost
and the $e^+e^-$ mass was reconstructed 0--0.5 MeV/$c^2$ high.

\begin{figure}
\includegraphics[width=3.5in]{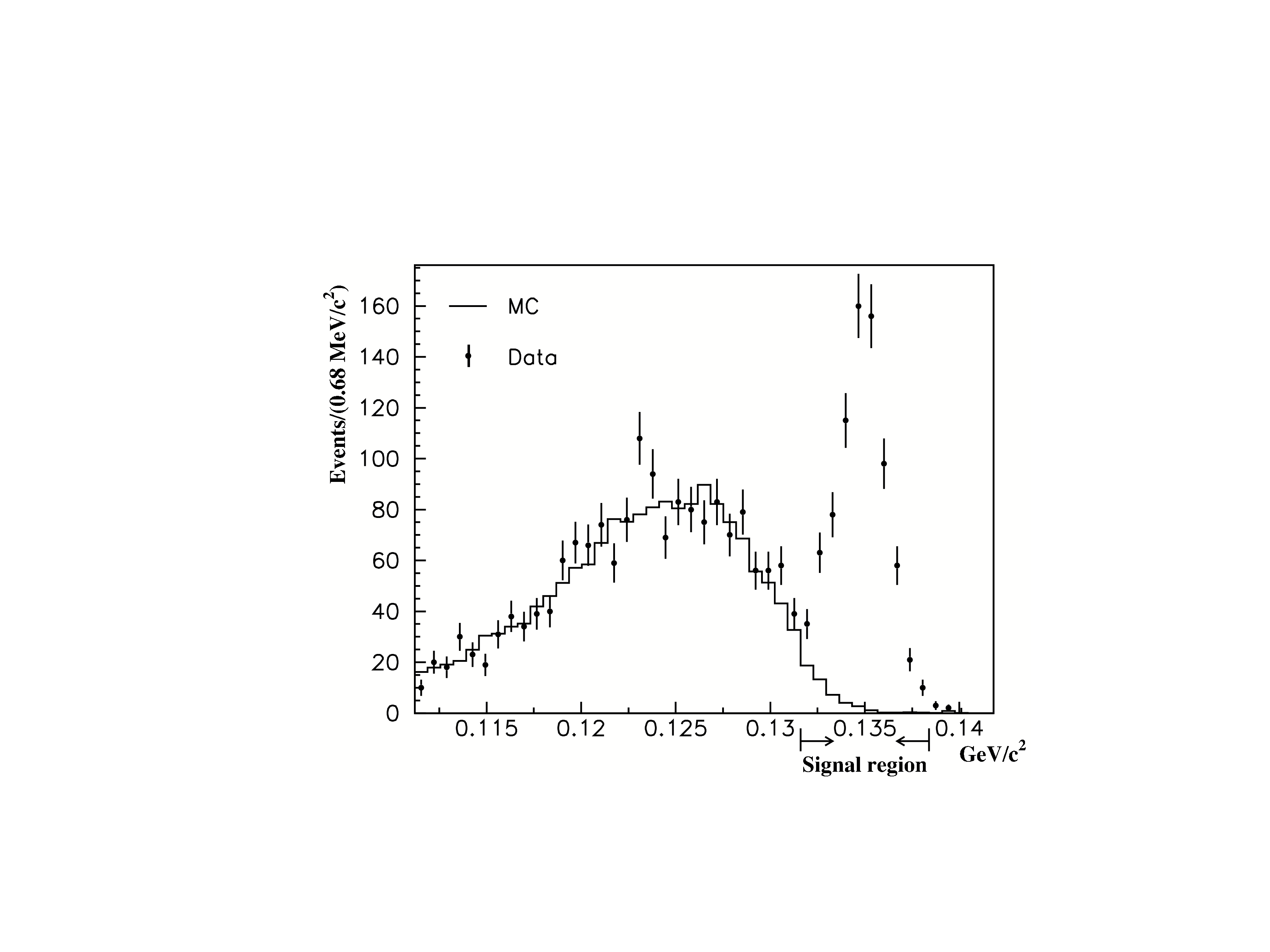}
\caption{\label{fig:mee}
Positron-electron invariant mass for \ptoee\ signal candidates passing
all other cuts.  The points
with error bars are data; the solid histogram is background MC.  }
\end{figure}

A plot of $m_{e^+e^-}$ after all cuts, Fig.~\ref{fig:mee}, shows the
signal peak at the $\pi^0$ mass and a background distribution that
extends under the peak. The background MC normalized by the measured
number of sideband Dalitz decays is plotted as well. The signal
region was $131.6<m_{e^+e^-}<138.4$~MeV/$c^2$, in which $794$ events
were found. The MC predicted a $2.94\%$ detector acceptance for the
signal in the 1997 run period and $3.14\%$ in 1999.  In the
normalization sample, $1~874~637$ candidates were found with 0.1\%
background. The acceptance for the normalization was $1.21\%$ in 1997
and $1.38\%$ in 1999.  The background in the signal region was
estimated using a MC simulation of each of the considered backgrounds.
Of these background events, 79\% were high $e^+e^-$-mass Dalitz
decays and the remainder were four-electron final states. 

\begin{table}
\begin{tabular}{| l | c |}
\hline
Branching ratio uncertainties & \\
\hline
\hline
 Statistical uncertainty & {\bf 3.8\%} \\
\hline
\ptoeeg\ branching ratio & 2.7\% \\
$\pi^0$ slope parameter & 1.3\% \\
 Total external systematic uncertainty & {\bf 3.0\%}\\
\hline
Background normalization & 1.1\% \\
$m_{e^+e^-}$ resolution & 0.7\% \\
Photon pairing $\chi^2$ modeling & 0.5\% \\
Kaon momentum spectrum & 0.4\% \\
$m_{e^+e^-}$ cutoff in normalization & 0.3\% \\
Background MC statistics & 0.4\% \\
Signal/normalization MC statistics & 0.3\% \\
Total internal systematic uncertainty & {\bf 1.6\%}\\
\hline
{\bf Total systematic uncertainty} & {\bf 3.4\%} \\
\hline
{\bf Total uncertainty on B(\ptoee )} & {\bf 5.1\%} \\
\hline
\end{tabular}
\caption{\label{tab:uncertainties} 
List of uncertainties in the \ptoee\ branching ratio.}
\end{table}

The important systematic error sources that were identified are listed
in Table \ref{tab:uncertainties}.  External systematic errors are
separated so the result may be corrected in the future if the
branching ratio of the Dalitz decay and the fraction of the decay in
the high-$x_D$ region of phase space are measured more precisely.  The
Dalitz branching ratio used was $B(\ptoeeg)=(1.198\pm 0.032)\%$ where
the relative error, 2.7\%, transfers directly into the
\ptoee\ branching ratio. The MC based on Ref.~\cite{mikaelian} was
used to determine the fraction of Dalitz events that had
$m_{e^+e^-}>65$ MeV/$c^2$, and this number depended on the $\pi^0$
form factor used. The result was $\Gamma(m_{e^+e^-}>65~{\rm
  MeV}/c^2)/\Gamma({\rm all\ Dalitz}) = 0.0319$ when using the 2004
PDG\cite{pdg} average for the $\pi^0$ form factor slope. The slope
value is dominated by a measurement in a region of spacelike momentum
transfer \cite{cello} where an extrapolation using vector meson
dominance was done. Our observed $m_{e^+e^-}$ distribution disagreed
with MC at the $1.8\sigma$ level and indicated
a value that would change the fraction of events in the
$m_{e^+e^-}>65$ MeV/$c^2$ tail by 1.3\%. This disagreement is quoted
here as a systematic error. The detector acceptance depended
negligibly on the form factor.

The remaining systematic errors were internal to the experiment.  The
combination of charged and neutral information in calculating the
decay vertex caused a small shift in the $m_{e^+e^-}$ distribution,
with the data moving by 0.2 MeV/$c^2$ more than the MC. The signal
region in data was shifted accordingly to compensate, and an
uncertainty in the signal acceptance and the background estimate was a
consequence. The shift changed the acceptance by 0.4\% and the
background estimate by 10.9\%.  The two errors combined into a 0.7\%
bias on the branching ratio, which was taken as a systematic error.

\begin{figure}
\includegraphics[width=3.5in]{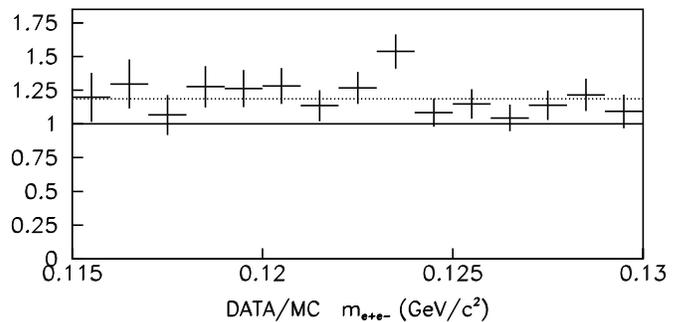}
\caption{\label{sideband-ratio}
Ratio of data to MC distribution of $m_{e^+e^-}$ in the
sideband region below the signal peak, where the MC was normalized
to the number of fully-reconstructed Dalitz decays. The dotted line indicates
the ratio over the entire region.}
\end{figure}

A systematic error was associated with the choice of normalization
for the background. Normalizing the prediction to the number of
fully-reconstructed Dalitz decays resulted in an estimate of $44.4\pm
2.7$ background events in the signal region, where the error is from
MC statistics only.  However, the data indicated a clear excess of
events in the sideband region, $110<m_{e^+e^-}<130$~MeV/$c^2$, over
this Dalitz-normalized MC.  The overall level
of background had to be scaled up by a factor of $1.19 \pm 0.04$ (1.24
in the 1997 data; 1.15 in the 1999 data) to match the data.
The relative excess showed little $m_{e^+e^-}$ dependence (see
Fig.~\ref{sideband-ratio}). There was no excess of events in the
sideband above the signal peak, which might have indicated an
unsimulated flat continuum background.  The source of the
low-$m_{e^+e^-}$ sideband excess was not fully understood, but was
likely related to modeling of the sensitivity of the veto system and
CsI to the soft photon from high-$x_D$ Dalitz decays.  The entire shift
was taken as a conservative systematic error.  This contributed a
1.1\% systematic uncertainty to the branching ratio.  The final
background estimate was $52.7\pm11.2$.

The high tail of the pairing $\chi^2$ distribution was not simulated
perfectly in the normalization and was a source of systematic
uncertainty. Removing the $\chi^2$ cut in the normalization analysis
changed the measured number of decaying kaons by 0.5\%.  This was not
expected to cancel in the ratio, as the pairing $\chi^2$ distributions
were different between signal and normalization due to the presence of
an additional photon in the normalization sample.  The entire
sensitivity of the normalization level to the cut was taken as a
systematic error. 

The simulated kaon momentum distribution deviated from the data,
as evidenced by a slope in the ratio of the reconstructed momentum
distributions in data and MC. Each MC event was reweighted to account
for the slope in both signal and normalization. This modification
changed the ratio of signal to normalization acceptances by 0.4\%,
which was taken as a systematic error on the branching ratio.
In the normalization, the cut on $m_{e^+e^-}$ caused a small bias in
the branching ratio due to modeling of the acceptance near the
$m_{e^+e^-}=70$ MeV/$c^2$ boundary. Tightening the cut by 5 MeV/$c^2$
produced a 0.4\% difference in the branching ratio.


\section{Results and conclusion}

The final branching ratio was calculated from 794 candidate signal
events with an estimated background of $52.7\pm11.2,$ and 
1~874~637 normalization events with negligible background. We found
\begin{equation}
\frac
{\Gamma(\ptoee,x_D>0.95)}
{\Gamma(\ptoeeg,x_D>0.232)} = (1.685 \pm 0.064 \pm 0.027)\times 10^{-4}
\end{equation}
where $x_D=0.232$ corresponds to $m_{e^+e^-} = 65$~MeV/$c^2$. Extrapolating the Dalitz branching
ratio to the full range of $x_D$ yields 
\begin{equation}
{\text{B}(\pi^0\rightarrow e^+e^-,~x_D>0.95)} = (6.44\pm0.25\pm0.22)\times 10^{-8}.
\end{equation}
In both cases the first error is from data statistics alone and the
second is the total systematic error.

Comparison with theoretical predictions and the unitary bound can be
done only if we remove the effects of final state radiation. This was done by
extrapolating the full radiative tail beyond $x_D=0.95$ and scaling the
result back up by the overall radiative correction of 3.4\% to find
the lowest-order rate for \ptoee .  We found
$\text{B}^{\text{no-rad}}(\ptoee)=(7.48\pm0.29\pm0.25)\times 10^{-8}$,
more than seven standard deviations higher than the unitary bound. The
result falls between VMD \cite{ametller} and $\chi$PT predictions
\cite{gomez}, with a significance on the difference of 2.3 and 1.5
standard deviations respectively.

\begin{acknowledgments}
We gratefully acknowledge the support and effort of the Fermilab
staff and the technical staffs of the participating institutions for
their vital contributions.  This work was supported in part by the U.S.
Department of Energy, The National Science Foundation, The Ministry of
Education and Science of Japan,
Funda\c{c}\~{a}o de Amparo a Pesquisa do Estado de S\~{a}o Paulo-FAPESP,
Conselho Nacional de Desenvolvimento Cientifico e Tecnologico-CNPq and
CAPES-Ministerio da Educa\c{c}\~{a}o.

\end{acknowledgments}

\bibliography{refs}

\end{document}